\documentclass[twocolumn]{aastex631}

\defcitealias{Gillanders24}{Gillanders, Rhodes \& Srivastav et al. 2024}
\defcitealias{LevanEP240314a}{Levan et al. 2024b}
\defcitealias{LiuEP240315a}{Liu et al. 2024c}

\newcommand{\eg}{\mbox{e.g.}}
\newcommand{\ie}{\mbox{i.e.}}

\newcommand\hostGalName{SDSS~J124601.99-094309.3}   
\newcommand\hostphotz{$0.299 \pm 0.043$}            
\newcommand\hostz{\mbox{$z = 0.4018 \pm 0.0010$}}   
\newcommand\hostDist{$2.25$\,{\rm Gpc}}             
\newcommand\hostDistMod{$41.76$}                    
\newcommand\hostAngSep{4\farcs9}                    
\newcommand\hostPhysicalSep{27\,{\rm kpc}}          

\newcommand\EPdiscoveryUTC{14 April 2024 at 09:50:12~UTC}  
\newcommand\EPdiscoveryMJD{60414.40986}                    

\newcommand{\thisEP}{\EPxx{240414a}}
\newcommand{\thisAT}{\ATxx{2024gsa}}

\newcommand{\SNxx}[1]{\mbox{SN\,#1}}
\newcommand{\ATxx}[1]{\mbox{AT\,#1}}
\newcommand{\GRBxx}[1]{\mbox{GRB\,#1}}
\newcommand{\EPxx}[1]{\mbox{EP#1}}

\newcommand{\I}{{\sc i}}
\newcommand{\II}{{\sc ii}}
\newcommand{\III}{{\sc iii}}

\newcommand{\V}{{\sc v}}

\newcommand{\ergscm}{erg\,s$^{-1}$\,cm$^{-2}$}
\newcommand{\ergs}{erg\,s$^{-1}$}
\newcommand{\msun}{M$_{\odot}$}

\newcommand{\mosfit}{\texttt{MOSFiT}}

\newcommand{\Swift}{\mbox{\textit{Swift}}}
\newcommand{\Chandra}{\mbox{\textit{Chandra}}}
\newcommand{\XMMNewton}{\mbox{\textit{XMM-Newton}}}

\usepackage{multirow}
\usepackage{booktabs}
\usepackage{subfigure}

\shorttitle{Optical counterpart discovery of EP240414a}
\shortauthors{Srivastav, Chen and Gillanders et al.}

\begin{document}

\title{Identification of the optical counterpart of the fast X-ray transient EP240414a}

\correspondingauthor{
    \\Shubham Srivastav (\href{mailto: shubham.srivastav@physics.ox.ac.uk}{shubham.srivastav@physics.ox.ac.uk}),
    \\T.-W. Chen (\href{mailto: twchen@astro.ncu.edu.tw}{twchen@astro.ncu.edu.tw}) \& 
    \\ James~H.~Gillanders (\href{mailto: james.gillanders@physics.ox.ac.uk}{james.gillanders@physics.ox.ac.uk}).
}

\author[0000-0003-4524-6883]{S.~Srivastav}
\thanks{These authors contributed equally.}
\affil{Astrophysics sub-Department, Department of Physics, University of Oxford, Keble Road, Oxford, OX1 3RH, UK}

\author[0000-0002-1066-6098]{T.-W.~Chen}
\thanks{These authors contributed equally.}
\affil{Graduate Institute of Astronomy, National Central University, 300 Jhongda Road, 32001 Jhongli, Taiwan}

\author[0000-0002-8094-6108]{J.~H.~Gillanders}
\thanks{These authors contributed equally.}
\affil{Astrophysics sub-Department, Department of Physics, University of Oxford, Keble Road, Oxford, OX1 3RH, UK}

\author[0000-0003-2705-4941]{L.~Rhodes}
\affil{Astrophysics sub-Department, Department of Physics, University of Oxford, Keble Road, Oxford, OX1 3RH, UK}
\affil{Trottier Space Institute at McGill, 3550 Rue University, Montreal, Quebec H3A 2A7, Canada}
\affil{Department of Physics, McGill University, 3600 Rue University, Montreal, Quebec H3A 2T8, Canada}

\author[0000-0002-8229-1731]{S.~J.~Smartt}
\affil{Astrophysics sub-Department, Department of Physics, University of Oxford, Keble Road, Oxford, OX1 3RH, UK}
\affil{Astrophysics Research Centre, School of Mathematics and Physics, Queen's University Belfast, BT7 1NN, UK}

\author[0000-0003-1059-9603]{M.~E.~Huber} 
\affiliation{Institute for Astronomy, University of Hawai'i, 2680 Woodlawn Drive, Honolulu, HI 96822, USA}

\author[0000-0002-9928-0369]{A.~Aryan} 
\affil{Graduate Institute of Astronomy, National Central University, 300 Jhongda Road, 32001 Jhongli, Taiwan}

\author[0000-0002-2898-6532]{S.~Yang} 
\affil{Henan Academy of Sciences, Zhengzhou 450046, Henan, China}

\author[0000-0003-3753-3102]{A.~Beri}
\affil{Indian Institute of Science Education and Research (IISER) Mohali, Punjab 140306, India}

\author[0000-0002-4033-3139]{A.~J.~Cooper}
\affil{Astrophysics sub-Department, Department of Physics, University of Oxford, Keble Road, Oxford, OX1 3RH, UK}

\author[0000-0002-2555-3192]{M.~Nicholl} 
\affil{Astrophysics Research Centre, School of Mathematics and Physics, Queen's University Belfast, BT7 1NN, UK}

\author[0000-0001-9535-3199]{K.~W.~Smith} 
\affil{Astrophysics Research Centre, School of Mathematics and Physics, Queen's University Belfast, BT7 1NN, UK}

\author[0000-0002-0504-4323]{H.~F.~Stevance}
\affil{Astrophysics sub-Department, Department of Physics, University of Oxford, Keble Road, Oxford, OX1 3RH, UK}
\affil{Astrophysics Research Centre, School of Mathematics and Physics, Queen's University Belfast, BT7 1NN, UK}

\author[0000-0002-0426-3276]{F.~Carotenuto}
\affil{Astrophysics sub-Department, Department of Physics, University of Oxford, Keble Road, Oxford, OX1 3RH, UK}

\author[0000-0001-6965-7789]{K.~C.~Chambers}
\affil{Institute for Astronomy, University of Hawai'i, 2680 Woodlawn Drive, Honolulu, HI 96822, USA}

\author[0000-0002-9085-8187]{A.~Aamer}
\affil{Astrophysics Research Centre, School of Mathematics and Physics, Queen's University Belfast, BT7 1NN, UK}

\author[0000-0002-4269-7999]{C.R.~Angus} 
\affil{Astrophysics Research Centre, School of Mathematics and Physics, Queen's University Belfast, BT7 1NN, UK}

\author[0000-0003-1916-0664]{M.~D.~Fulton} 
\affil{Astrophysics Research Centre, School of Mathematics and Physics, Queen's University Belfast, BT7 1NN, UK}

\author[0000-0001-8385-3727]{T.~Moore} 
\affil{Astrophysics Research Centre, School of Mathematics and Physics, Queen's University Belfast, BT7 1NN, UK}

\author[0000-0001-8605-5608]{I. A. Smith}
\affiliation{Institute for Astronomy, University of Hawai'i, 34 Ohia Ku St., Pukalani, HI 96768-8288, USA}

\author[0000-0002-1229-2499]{D.~R.~Young}
\affil{Astrophysics Research Centre, School of Mathematics and Physics, Queen's University Belfast, BT7 1NN, UK}

\author[0000-0001-5486-2747]{T.~de~Boer} 
\affil{Institute for Astronomy, University of Hawai'i, 2680 Woodlawn Drive, Honolulu, HI 96822, USA}

\author[0000-0003-1015-5367]{H.~Gao} 
\affil{Institute for Astronomy, University of Hawai'i, 2680 Woodlawn Drive, Honolulu, HI 96822, USA}

\author[0000-0002-7272-5129]{C.-C.~Lin}
\affil{Institute for Astronomy, University of Hawai'i, 2680 Woodlawn Drive, Honolulu, HI 96822, USA}

\author[0000-0002-9438-3617]{T.~Lowe}
\affil{Institute for Astronomy, University of Hawai'i, 2680 Woodlawn Drive, Honolulu, HI 96822, USA}

\author[0000-0002-7965-2815]{E.~A.~Magnier} 
\affil{Institute for Astronomy, University of Hawai'i, 2680 Woodlawn Drive, Honolulu, HI 96822, USA}

\author{P.~Minguez}
\affil{Institute for Astronomy, University of Hawai'i, 2680 Woodlawn Drive, Honolulu, HI 96822, USA}

\author[0000-0001-8415-6720]{Y.-C.~Pan} 
\affil{Graduate Institute of Astronomy, National Central University, 300 Jhongda Road, 32001 Jhongli, Taiwan}

\author[0000-0002-1341-0952]{R.~J.~Wainscoat}
\affil{Institute for Astronomy, University of Hawai'i, 2680 Woodlawn Drive, Honolulu, HI 96822, USA}

\begin{abstract}
    \noindent
    Fast X-ray transients (FXTs) are extragalactic bursts of X-rays first identified in archival X-ray data, and now routinely discovered by the Einstein Probe in real time, which is continuously surveying the night sky in the soft ($0.5 - 4$\,keV) X-ray regime. In this Letter, we report the discovery of the second optical counterpart (\thisAT) to an FXT (\thisEP). \thisEP\ is located at a projected radial separation of 27\,kpc from its likely host galaxy at $z = 0.4018 \pm 0.0010$. The optical light curve of \thisAT\ displays three distinct components. The initial decay from our first observation is followed by a re-brightening episode, displaying a rapid rise in luminosity to an absolute magnitude of $M_r \sim -21$ after two rest-frame days. While the early optical luminosity and decline rate is similar to luminous fast blue optical transients, the colour temperature of \thisAT\ is distinctly red and we show that the peak flux is inconsistent with a thermal origin. The third component peaks at $M_i \sim -19$ at $\gtrsim 16$ rest-frame days post-FXT, and is compatible with an emerging supernova. We fit the $riz$-band data with a series of power laws and find that the decaying components are in agreement with gamma-ray burst afterglow models, and that the re-brightening may originate from refreshed shocks. By considering \thisEP\ in context with all previously reported known-redshift FXT events, we propose that Einstein Probe FXT discoveries may predominantly result from (high-redshift) gamma-ray bursts, and thus appear to be distinct from the previously discovered lower redshift, lower luminosity population of FXTs. 
\end{abstract}

\keywords{
    Transient sources (1851);
    Relativistic jets (1390);
    High-energy astrophysics (739);
    X-ray transient sources (1852);
    Optical identification (1167)
}

\section{Introduction} \label{sec:Introduction}

Fast X-ray transients (FXTs) are extragalactic, short duration bursts of X-rays lasting a few seconds to thousands of seconds,
typically not associated with a higher (\ie, gamma-ray) or lower (\ie, optical, radio) energy counterpart. The field of FXTs is undergoing a revolution since the launch and operation of the Einstein Probe \citep[EP;][]{Yuan2022_EinsteinProbe} in early 2024. The wide field of view (3600\,deg$^2$ instantaneous) allows the sky to be surveyed multiple times in a 24\,hr period with the wide-field X-ray Telescope onboard the Einstein Probe (EP-WXT). Its lobster-eye technology typically provides a localisation of 3~arcmin radius in the soft X-ray band of $0.5 - 4$\,keV \citep[\eg,][]{ATel16463, Zhang2024_GCN35931}, with a sensitivity of $1.2 \times 10^{-10}$\,\ergscm\ for an exposure of 100\,s \citep{Yuan2022_EinsteinProbe}. The survey power is illustrated by the grasp of the EP-WXT (effective area~$\times$~field-of-view) being a factor of 10 greater than \XMMNewton, \textit{ROSAT} and \textit{eROSITA} \citep[see figure~10 of][]{Yuan2022_EinsteinProbe}. EP's follow-up X-ray telescope (EP-FXT) can be triggered to follow up and improve the localisation of sources from the EP-WXT to a much smaller radius of about 10~arcsec, while also being significantly more sensitive. 

Einstein Probe has released 27 discoveries\footnote{As of 14 August 2024.} of candidate extragalactic FXTs through GCNs as well as a number of likely Galactic sources such as EPW\,20240305aa \citep[][spatially consistent with a stellar source]{ATel16509, ATel16514, ATel16529} and \EPxx{240309a} \citep[][a known variable X-ray source]{ATel16546}. To uncover the nature of these FXTs, the key piece of information required is the luminosity distance, which requires the detection of an optical counterpart and a spectrum to determine redshift. Of the 27 extragalactic FXTs reported via GCNs, optical counterparts have been detected for eight, and redshifts have been measured for five. The first counterpart discovered was that of \EPxx{240315a} \citepalias[\ATxx{2024eju};][]{Gillanders24}, which had a remarkably high redshift of $z = 4.859$ \citepalias{Gillanders24, LevanEP240314a, LiuEP240315a} and was associated with \GRBxx{240315C} \citep{2024GCN.35971....1D, LiuEP240315a, 2024GCN.35972....1S}. Seven of these 27 FXTs have a spatial and temporal coincidence with a GRB. The detections and limits obtained thus far indicate that FXT counterparts are typically fainter than $20 - 21$~mag. Apart from the high redshift of \EPxx{240315a}/\ATxx{2024eju}/\GRBxx{240315C}, the other FXTs with redshifts lie between $0.4 \lesssim z \lesssim 3.6$ \citep{JonkerGCN36110, QuirolaVasquezGCN37013, QuirolaVasquezGCN37087}, with the next highest being \EPxx{240804a} at $z = 3.662$ \citep{2024GCN.37039....1B}. This may indicate that the majority of EP FXTs discovered to date, and potentially most that will be discovered by the mission, are bright X-ray counterparts to high-redshift GRBs, and thus EP observations may signpost a method to locate distant explosions for rapid follow-up observations. The 3~arcmin error radius provided by EP-WXT enables many facilities to quickly and sensitively search for optical counterparts.

Before the launch of EP, FXTs had been discovered in archival images from the more sensitive and narrower field-of-view missions \textit{Chandra} and \textit{XMM-Newton} \citep{jonker13,glennie15,bauer17}. 12 FXTs uncovered in \textit{XMM-Newton} data were reported by \cite{alp_larssson_20}, and  \cite{eappachen24} showed the spectroscopic redshifts of seven plausible hosts lie in the range, $0.098< z < 0.645$. Supernova shock breakout (SBO) from blue and red supergiants, or from Wolf-Rayet progenitors surrounded by a dense circumstellar medium, has been suggested as the origin for these archival FXTs by \cite{alp_larssson_20}, although the higher redshifts of the putative hosts now challenge that interpretation. Other explanations proposed for these FXTs are binary neutron star (BNS) mergers producing a magnetar central engine \citep{xue19, ai_zhang_21, sarin21,  Eappachen2023a, QuirolaVasquez24} and the tidal disruption of white dwarfs by intermediate-mass black holes \citep{macleod16}.  

The extensive search through over 22 years of \Chandra\ archival data by \cite{Quirola2022,Quirola2023} is now providing sky rates and quantifying the X-ray fluxes and durations of FXTs. Given that EP FXTs are typically $100 - 1000$ times brighter than the \Chandra\ and \XMMNewton\ sources, the link between the two populations is an open question. The large X-ray fluxes, high redshifts and GRB coincidences of the EP-detected FXTs to date are in contrast to the fainter and lower redshift \Chandra\ and \XMMNewton\ objects. 

In this Letter, we present the optical counterpart discovery and follow-up to the FXT \thisEP, which, to date, has the lowest redshift of any counterpart to an extragalactic EP source. As such, it may provide insight into the links between these two populations. \thisEP\ was detected on \EPdiscoveryUTC\ (MJD~\EPdiscoveryMJD) by the wide-field X-ray telescope with a peak flux, \mbox{$f_{\rm X} \sim 3 \times 10^{-9}$\,\ergscm} in the $0.5 - 4$\,keV band \citep{LianWXT_GCN36091}. A counterpart and host galaxy were quickly reported \citep{AryanGCN36094}, followed by a redshift of the host \citep[$z \simeq 0.41$;][]{JonkerGCN36110}. Remarkably, the optical counterpart was observed to re-brighten \citep{SrivastavGCN36150}, a signature of a supernova (SN) was reported \citep{LevanGCN36355}, and a luminous radio counterpart was also reported \citep{BrightGCN36362}. 

Throughout this paper we assume $\Lambda$CDM cosmology with a Hubble constant, $H_0 = 67.7$\,km\,s$^{-1}$\,Mpc$^{-1}$, $\Omega_{\rm M} = 0.309$ and \mbox{$\Omega_{\Lambda} = 0.691$} \citep{2016A&A...594A..13P}. We also assume a line-of-sight Milky Way extinction of \mbox{$E(B-V) = 0.033$} mag \citep{Schlafly2011}.

\begin{figure*}
    \centering
    \includegraphics[width=0.8\linewidth]{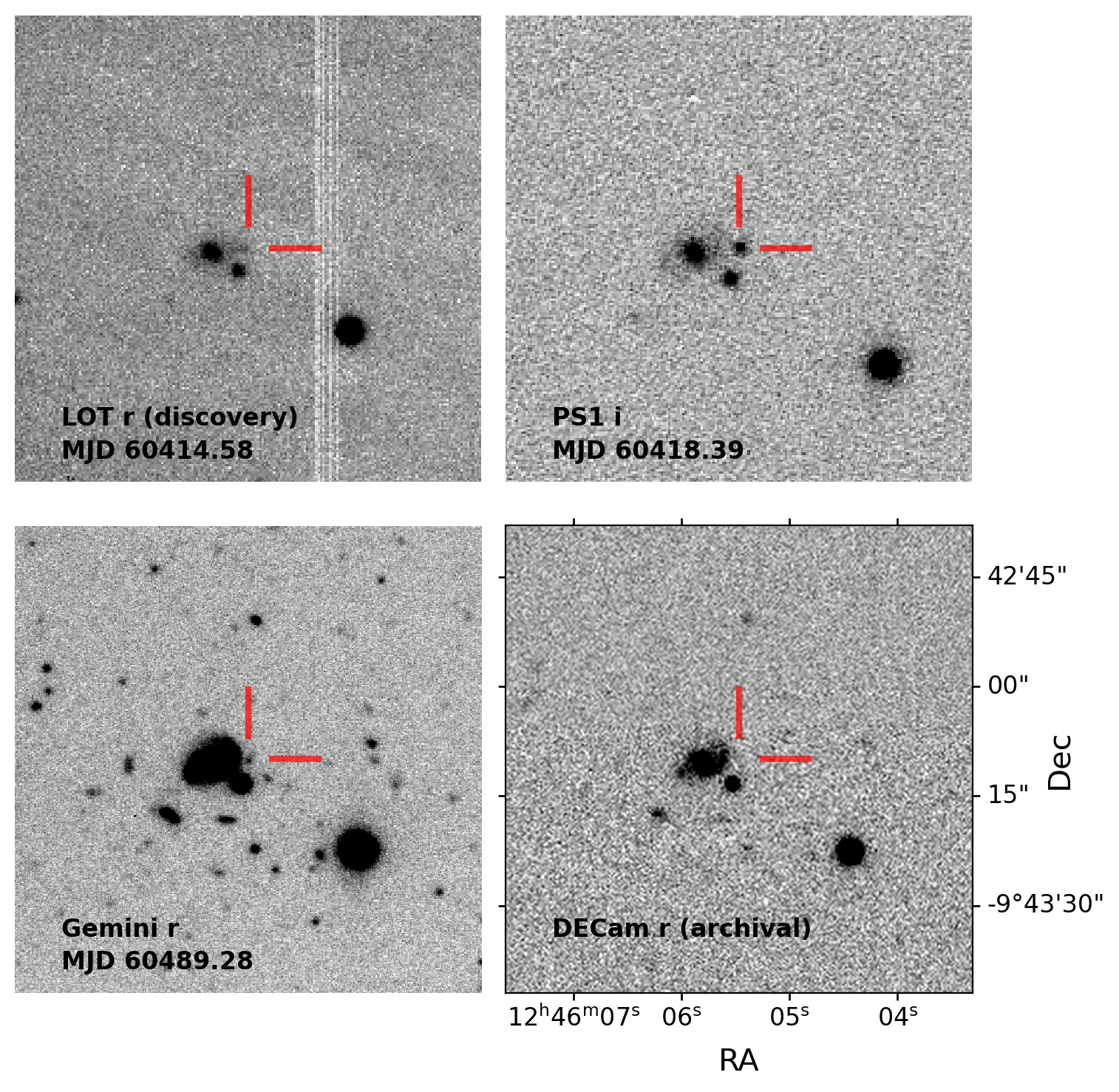}
    \caption{
        LOT discovery image of \thisAT, along with follow-up images obtained with PS1 and Gemini. Also shown (bottom right) is an archival image of the field from the Dark Energy Survey Camera (DECam).
    }
    \label{fig:AT204gsa detection images}
\end{figure*}

\section{Discovery of the optical counterpart with LOT} \label{sec:DiscoveryLOT}

The 3~arcmin uncertainty in the position of \thisEP\ and the discovery time provided by EP-WXT \citep{EP240414aDiscovery} offered the opportunity for the telescopes at the Lulin Observatory in Taiwan to search for a counterpart within the first few hours of the FXT alert. As part of our ongoing program on fast response follow-up for multi-wavelength and multi-messenger transients at Lulin \citep[Kinder; see][]{Chen2021_KINDER, Chen2024}, we observed the field of \thisEP\ using the Lulin One-meter Telescope (LOT). The LOT is equipped with a $2048 \times 2048$~pixel CCD, providing a 12~arcmin field-of-view, and a set of SDSS $u'g'r'i'z'$ filters. The first LOT epoch of observations started on 14~April~2024 at 12:58~UTC (MJD 60414.540), 3.12\,hrs following the EP trigger. We obtained a sequence of $6 \times 300$\,s images, alternating between the $r$ and $i$ filters. Approximately three hours later, we conducted observations in the $g$-band, following the same $6 \times 300$\,s sequence. The transient was clearly identified by eye in the first set of co-added $r$-band images.

The data were processed in a standard manner; images were stacked and template images from the Legacy Survey DR10 \citep{LegacySurveys} were used as reference to produce difference images for rapid transient searches \citep[see][for a description of the pipeline]{Yang21}. The new optical transient, \thisAT, was discovered in the difference images \citep{EP240414aKinderDiscovery}, and was reported to the Transient Name Server \citep{AT2024gsaTNSreport} with coordinates, \mbox{${\rm RA} = +12^{\rm h} 46^{\rm m} 01.67^{\rm s}$}, \mbox{${\rm Dec} = -09^\circ 43' 08\farcs8$}. We measured discovery magnitudes of $r = 21.52 \pm 0.12$ and $i = 21.40 \pm 0.16$ (AB mag). The two nearest catalogued optical sources in the Sloan Digital Sky Survey (SDSS) DR15 \citep{SDSSDR15}, the Pan-STARRS1 $3 \pi$ survey DR2 \citep{flewelling2020}, and the DESI Legacy Surveys DR10 \citep{LegacySurveys} are 3\farcs6 and 4\farcs9 offset from \thisAT. The nearest source, SDSS~J124601.74-094312.1 is morphologically classified as a stellar object ($r = 20.27 \pm 0.03$ AB mag) in both SDSS DR15 and the Legacy Survey DR10. The most likely host is the brighter \mbox{($r = 19.04 \pm 0.02$~AB~mag)} galaxy to the east, SDSS~J124601.99-094309.3, with coordinates, ${\rm RA} = 12^{\rm h} 46^{\rm m} 01.99^{\rm s}$, ${\rm Dec} = -09^\circ 43' 09\farcs 34$. \thisAT\ has a projected 4\farcs9 offset from this galaxy, which has a photometric redshift of $z=$~\hostphotz\ from SDSS DR15. The discovery images and the host galaxy are shown in Figure~\ref{fig:AT204gsa detection images}.

This source was confirmed as the likely counterpart with the 2.56-m Nordic Optical Telescope at 12.29\,hrs after the EP trigger \citep{EP240414aNOTXu}, and a spectrum taken at approximately +15\,hrs by \cite{EP240414aGTC} indicated a featureless continuum and a redshift of \mbox{$z = 0.41$} for the most likely host galaxy SDSS~J124601.99-094309.3.

Following discovery, we launched a multi-wavelength observing campaign to sample the evolution of the optical counterpart; for details of our observing campaign, see Appendix~\ref{SEC: Appendix - Optical counterpart follow-up}. Our optical light curve is presented in Figure~\ref{fig:AT2024gsa_LC} and the summary of photometric observations is provided in Table~\ref{tab:Photometry}. We also present the \Swift /XRT data and use it to constrain the optical to X-ray spectral energy distribution.

\begin{figure*}
    \centering
    \includegraphics[width=\linewidth]{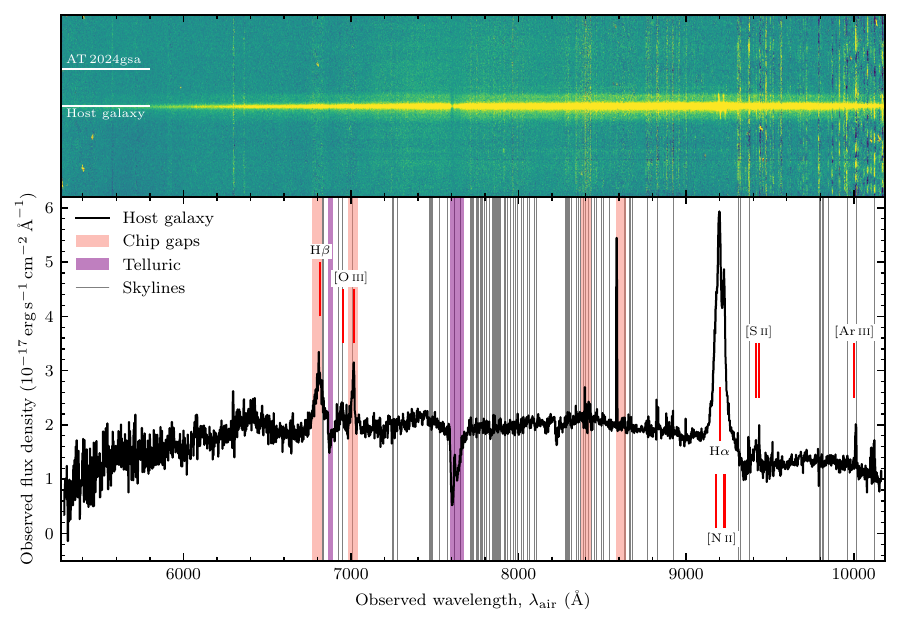}
    \caption{
        \textit{Upper panel:} 2D Gemini-South GMOS-S spectrum of \thisAT\ and its host galaxy, \hostGalName. The location of the host galaxy and transient traces are marked.
        \textit{Lower panel:} 1D spectrum of the host galaxy of \thisAT. The prominent emission features from which we extract a redshift estimate are marked and labelled. Regions corresponding to chip gaps and telluric absorption are shaded, and prominent skylines \citep[from][]{Hanuschik2003} are overlaid.
    }
    \label{fig:Gemini host spectrum}
\end{figure*}

\setlength{\abovecaptionskip}{-40pt} 
\begin{figure*}
    \centering
    \includegraphics[width=0.95\linewidth]{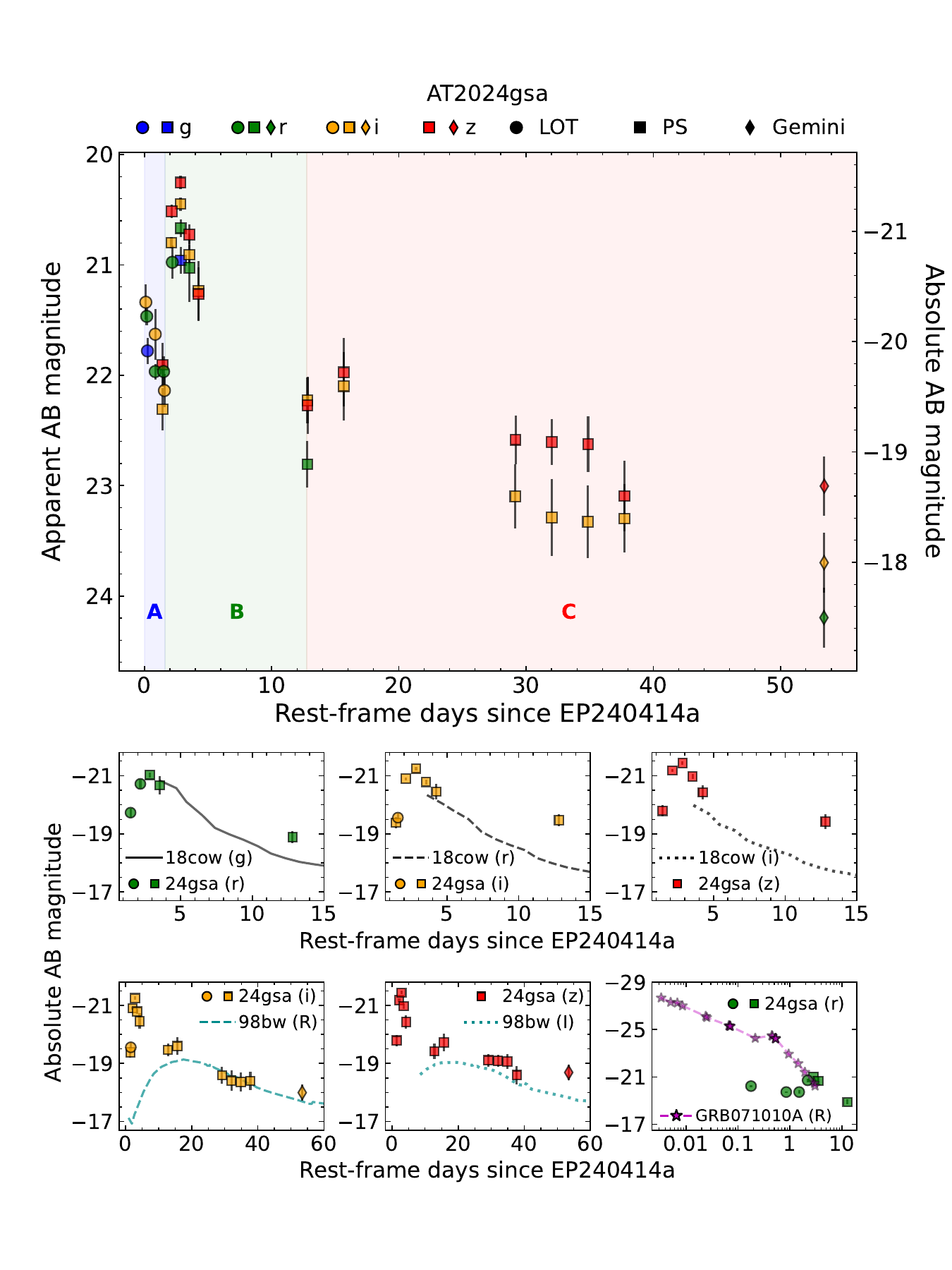}
    \caption{
        \textit{Upper panel:} Multi-band optical ($griz$) light curves of \thisAT\ obtained with LOT, PS and Gemini. Both apparent and absolute magnitude scales are shown. The shaded regions highlight the three distinct phases in the optical light curve.
        \textit{Middle panels:} Comparison of the rapidly evolving phase B of \thisAT\ with the well-studied FBOT \ATxx{2018cow} \citep{2018ApJ...865L...3P}.
        \textit{Lower left and centre panels:} Comparison of the slowly evolving phase C of \thisAT\ with the broad-line Ic \SNxx{1998bw} \citep{2011AJ....141..163C}.
        \textit{Lower right panel:} Comparison of \thisAT\ ($r$-band) at early times with the $R$-band light curve of \GRBxx{071010A}, which exhibited a re-brightening episode $\approx 0.6$\,d post-GRB \citep{2008MNRAS.388..347C}.
    }
    \label{fig:AT2024gsa_LC}
\end{figure*}
\setlength{\abovecaptionskip}{10pt} 

\section{Spectroscopic observation with Gemini and host redshift measurement} \label{SEC: Redshift estimation}

With a Gemini-South Director's Discretionary Time proposal (PI: S.~J.~Smartt) we were awarded time for longslit spectroscopy of the transient and its host galaxy. However, due to weather and instrument availability constraints we could only observe once the transient had faded from detectability. We centred the slit of the Gemini Multi-Object Spectrograph (GMOS-S) on the host galaxy \hostGalName, with a slit angle of 95.2$^{\circ}$ East of North. This was to capture the position of \thisAT\ at a separation of \hostAngSep\ (see Figure~\ref{fig:AT204gsa detection images}) from the host. We obtained $4 \times 1800$\,s exposures on 17 July 2024 (commencing at 23:17:27~UT/MJD~60508.970) with the R400 grism, the GG455 blocking filter and 1\farcs0 slit width. Dithering was employed in both the spatial (15\farcs0, for sky subtraction) and wavelength (200\,\AA, to cover the chip gaps) directions.

We used the \texttt{DRAGONS} pipeline \citep{Labrie2023_DRAGONS, DRAGONS_zenodo} to reduce the data following standard recipes, with the reduced spectrum flux-calibrated against a standard star. This produced the combined, stacked spectrum shown in Figure~\ref{fig:Gemini host spectrum}. Strong emission lines of H$\alpha$, H$\beta$, [N\,\II]~$\lambda \lambda 6548.050, 6583.460$, [S\,\II]~$\lambda \lambda 6716.440, 6730.810$, [O\,\III]~$\lambda \lambda 4958.911, 5006.843$ and [Ar\,\III]~$\lambda 7135.790$ are observed at a common redshift of \hostz. In comparison, the photometric redshift of the host from SDSS DR15 was estimated to be $z_{\rm phot}=~$\hostphotz\ \citep{SDSSDR15}. At a redshift of $z = 0.4018$, the luminosity distance and distance modulus derived assuming $\Lambda$CDM cosmology \citep[as outlined in Section~\ref{sec:Introduction}; see also][]{Planck2016Cosmo} are $D_{\rm L} =$~\hostDist\ and $\mu =$~\hostDistMod, respectively. The angular scale is 5.55\,kpc\,arcsec$^{-1}$, implying a projected physical separation of $R_g =$~\hostPhysicalSep\ between the host galaxy and the position of \thisAT. The slit position of GMOS-S was orientated to include the position of \thisAT\ at a separation of \hostAngSep. There is no detection at any continuum or emission line flux at the position of \thisEP/\thisAT\ (see Figure~\ref{fig:Gemini host spectrum}).

Finally, we estimate the chance coincidence of \thisEP/\thisAT\ with \hostGalName\ using the procedure outlined by \cite{Bloom2002}. We find that the probability of chance coincidence is $\lesssim 0.01$, indicating that the transient is almost certainly located at $z = 0.4018$, and is indeed associated with \hostGalName\
\citep[see also][for additional observational evidence in favour of a $z \simeq 0.40$ location]{vanDalenEP240414a}.

\section{Light curve analysis} \label{sec:analysis}

The optical light curve of \thisAT\ is shown in Figure~\ref{fig:AT2024gsa_LC}. The shaded regions highlight distinct phases in the light curve evolution. Phase~A represents the rapidly declining early light curve lasting $\sim 1.5$ rest-frame days post-FXT. \thisAT\ was discovered a few hours following the FXT (see Section~\ref{sec:DiscoveryLOT}) with a luminosity of $M_r \approx -20$ (AB mag).

We consider the possibility of phase A resulting from a cooling tail following shock breakout from an extended stellar envelope \citep{2017hsn..book..967W}, although this interpretation is not supported by the high luminosity and red colour ($g-i \approx 0.4$; observer frame). Moreover, the $0.4 - 4$\,keV X-ray flux reported by EP \citep{EP240414aDiscovery} implies an X-ray luminosity of $\sim 10^{48}$\,\ergs, 2 orders of magnitude higher than the candidate SN shock breakout events identified by \citet{alp_larssson_20}. Since shock breakout emission is thermal in nature, the combination of high X-ray luminosity and red optical colour is inconsistent with this interpretation.

Phase B (spanning $\sim 1.5 - 10$ rest-frame days) highlights an abrupt, rapid and achromatic rise in flux by a factor of $\gtrsim 3$, followed by a fast decline. The rapid timescale ($\sim 2$\,d rise) and absolute magnitude ($\sim -21$~AB~mag) during the peak of phase~B bears resemblance to the observed characteristics of luminous fast blue optical transients \citep[LFBOTs; \eg,][]{2019MNRAS.484.1031P}. However, the red colour of \thisAT\ during phase~B's peak ($r-z \approx 0.4 \pm 0.1$; observer frame) is at odds with LFBOTs. The middle panels in Figure~\ref{fig:AT2024gsa_LC} show light curve comparisons between \ATxx{2018cow} \citep{2018ApJ...865L...3P} and \thisAT. Given the redshift of the likely host galaxy for \thisAT\ ($z = 0.4018$; see Section~\ref{SEC: Redshift estimation}), we plot the observed $riz$ light curves of \thisAT\ alongside the $gri$ light curves of \ATxx{2018cow}, respectively, for a more meaningful comparison. Although the timescales for the $riz$ ($gri$) light curves for \thisAT\ (\ATxx{2018cow}) are similar, \thisAT\ is significantly more luminous at peak in the redder $iz$ bands, highlighting the intrinsically red colour in stark contrast to \ATxx{2018cow} and LFBOTs in general.

Assuming the phase B peak is thermally driven, the absolute magnitude of $M_r \approx -21$ would imply a luminosity of $L \simeq 8 \times 10^{43}\,$\ergs\ (with zero bolometric correction). We fit the $riz$-band photometry during phase~B with the \texttt{SuperBol} code \citep{2018RNAAS...2..230N}. \texttt{SuperBol} converts the extinction-corrected broadband photometry into monochromatic fluxes and performs a blackbody fit (redshift effects are accounted for). We infer a blackbody temperature, $T_{\rm bb} \approx 8000 \pm 2000$\,K for \thisAT\ around phase B peak. The large uncertainty on the temperature is due to the fact that the blackbody fit is performed on only three optical bands. The inferred luminosity around phase B peak, integrated across just the $gri$-bands, is $L_{gri} \approx 2.5 \times 10^{43}$\,\ergs, whereas the total luminosity integrated over the extrapolated blackbody fit is $L_{\rm bb} \approx 8.5 \times 10^{43}$\,\ergs. The fit yields a photospheric radius, $R_{\rm bb} \approx 5 \times 10^{15}$\,cm. This radius is very large for a compact progenitor origin, thus requiring an extreme ejecta velocity ($v_{\rm ej} \sim 0.7$\,c), given the rapid timescale of $\sim 3$~days to (phase~B) peak. Although the radius may be plausible for an extended stellar envelope or CSM distribution, as alluded to earlier, the combination of luminous X-ray emission and red optical colours seem to rule out a thermal emission mechanism. This is further illustrated by the requirement of $E_{\rm kin} \gtrsim 4 \times 10^{52}$\,erg to eject just 0.1\,\msun\ at this velocity. Thermal emission models for \ATxx{2018cow} and other LFBOTs can reproduce the observed light curves as they are extremely hot ($T_{\rm bb} \simeq 30$,000\,K) at peak. The red colour of \thisAT\ makes this explanation untenable for the observed peak at $\sim 3$~days. 
We present a color comparison for \thisAT\ with \ATxx{2018cow}, \GRBxx{071010A} and \SNxx{1998bw} in Appendix~\ref{sec:color_evol} and Figure~\ref{fig:color}. This illustrates the red color of AT2024gsa, and while the early color evolution may be similar to \SNxx{1998bw}, the very high luminosity requires a large photospheric radius and thus an extreme ejecta velocity, as described above.

Based on these simple calculations, we undertake a more robust analysis of the phase B light curve of \thisAT\ using the publicly available Modular Open Source Fitter for Transients \citep[\mosfit;][]{2017ApJ...850...55N,2018ApJS..236....6G}. We attempt to fit three models -- the \texttt{default} model powered by the radioactive decay of $^{56}$Ni \citep{1994ApJS...92..527N}, the \texttt{magni} model combining the luminosity from $^{56}$Ni decay and energy injection from the spin-down of a rapidly rotating magnetar \citep{2010ApJ...717..245K}, and the \texttt{magnetar} model powered only by the magnetar component. The \texttt{default} model produces the poorest fit, which is expected given the combination of high luminosity and rapid timescale involved. Although the magnetar-driven models yield better fits to the light curve, all models require extreme ejecta velocities in the range $0.8 \lesssim v_{\rm ej} / {\rm c} \lesssim 0.9$. As highlighted in the order-of-magnitude calculations above, this is owing to the intrinsically red colour of \thisAT\ that forces the model to converge to a low effective temperature. The high luminosity can thus only be explained by a large photospheric radius, thereby leading to very high expansion velocities. Based on these arguments, we deem these models to be unphysical. The overall characteristics of the light curve, including the high luminosity, rapid evolution, red colour and achromaticity suggest a non-thermal powering mechanism instead as the most likely explanation.

In the bottom-right panel of Figure~\ref{fig:AT2024gsa_LC}, we compare the $r$-band light curve of \thisAT\ with $R$-band observations of the afterglow of \GRBxx{071010A} \citep{2008MNRAS.388..347C}. The \GRBxx{071010A} afterglow exhibited an achromatic and abrupt re-brightening episode $\approx 0.6$~rest-frame days post-GRB, slightly earlier than the re-brightening observed in \thisAT. Late-time re-brightening episodes such as the one observed in \GRBxx{071010A} are not unheard of in GRB afterglow light curves, and are often attributed to additional energy injection from refreshed shocks \citep{rees_refreshed_1998}. We discuss this interpretation in more detail in Section~\ref{subsec:nature}. 

Finally, the third phase that we define for the light curve (phase C) appears to show marginal evidence for a rising component after $\sim 12$~rest-frame days, although our photometric coverage at these epochs was affected by the Moon and adverse weather. Nevertheless, it is clear that the rapid initial evolution during phases A and B has slowed down at these later epochs. The timescale and absolute magnitude for the $iz$-band light curves of \thisAT\ at this phase are compatible with an emerging SN component. Broad-line Ic SNe are often found in association with long GRBs. In the bottom panels of Figure~\ref{fig:AT2024gsa_LC}, we compare the $iz$ light curve of \thisAT\ with the $RI$ light curves of the broad-line Ic \SNxx{1998bw} \citep{2011AJ....141..163C} associated with \GRBxx{980425} \citep{1999ApJ...516..788W}. The comparison shows that the presence of an underlying SN component is plausible, supporting a long-duration GRB interpretation for \thisEP\ \citep[see also][]{BrightEP240414a,vanDalenEP240414a}. However, the lack of multi-band, high-cadence photometric coverage during phase C limits any definitive conclusions.

\section{Discussion} \label{sec:Discussion}

\subsection{The nature of \thisEP/\thisAT} \label{subsec:nature}

Given the findings and conclusions drawn in Section~\ref{sec:analysis}, we examine the optical light curves and spectra of \thisAT\ in the context of external shock models. 

The most striking signature of the optical light curves is the very rapid rise between $1.5 - 3$~rest-frame days post-FXT. For the bands where we have the most comprehensive coverage ($riz$), we connect the points either side of the sharp rise between 2 and 3 days with a simple power law model (no observations were made during the rise). We measure temporal indices of $2.4 \pm 0.5$, $4.1 \pm 0.5$ and $3.1 \pm 0.5$ for the $r$, $i$ and $z$ bands, respectively. 

\begin{figure}
    \centering
    \includegraphics[width=\columnwidth]{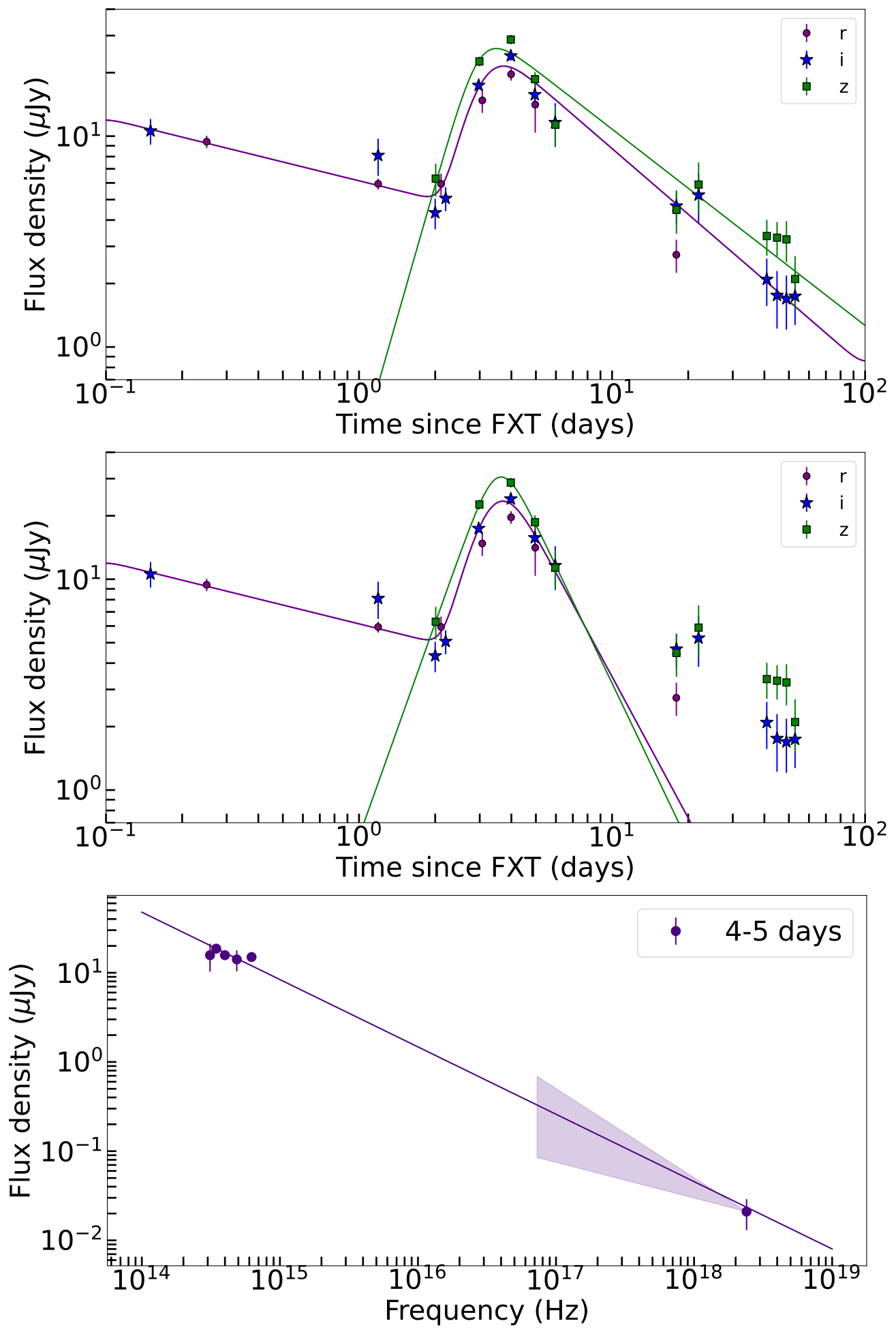}
    \caption{
        \textit{Top panel}: Power law $+$ broken power law fits to the full \mbox{$riz$-band} data for \thisAT. Temporal indices during the initial decay around phase A are \mbox{$\alpha_1 = -0.31 \pm 0.02$}, $-0.33 \pm 0.05$ for $r$ and $i$-band respectively. The indices for the post-peak decline are $\alpha_3= -1.4 \pm 0.4$, $-1.04 \pm 0.07$ and $-0.93 \pm 0.09$ for $r$, $i$ and $z$-band respectively.} \textit{Middle panel}: Power law $+$ broken power law fits to the $< 10$\,d \mbox{$riz$-band} data of \thisAT. The indices for the post-peak decline are steeper after excluding the phase C component, with $\alpha_3= -2.6 \pm 0.4$, $-2.2 \pm 0.4$ and $-2.5 \pm 0.1$ for $r$, $i$ and $z$-band, respectively. \textit{Bottom panel}: Optical-to-X-ray spectrum of the broadband counterpart of \EPxx{240414a}, containing data obtained between $4 - 5$~days post-FXT trigger.
    
    \label{fig:powerlaw_fits}
\end{figure}

We search for other possible frequency-dependent light curve behaviour by fitting the $ri$-band data with a power law to the initial decay phase (\ie, phase~A) of the form:
\begin{equation}
     F_{\nu} = A t^{\alpha_{1}},
     \label{EQN: PL}
\end{equation}
followed by a broken power law given by:
\begin{equation}
    F_{\nu} = B \left[ 0.5 \left( \frac{t}{t_{b}} \right)^{s \alpha_2} + 0.5 \left( \frac{t}{t_{b}} \right)^{s \alpha_3} \right]^{1 / s},
    \label{EQN:broken-PL}
\end{equation}
where $F_\nu$ is the flux density, $t$ is the time since FXT, $A$ and $B$ are  the amplitudes of the single and broken power law components, respectively, $t_b$ is the break time of the broken power law, and $\alpha_1$, $\alpha_2$, $\alpha_3$ are the different temporal power law exponents. We fit Equation~\ref{EQN:broken-PL} to the $z$-band data, and the resulting fits are shown in Figure~\ref{fig:powerlaw_fits}. The upper panel shows the fits to all the data, while the middle panel shows the fits to all the data obtained $< 10$~days post-FXT (to account for phase C emission likely being dominated by radiation from the SN).

For all epochs where we have data in at least three bands, we estimate the spectral index, assuming a single-component power law can describe the spectral index. We obtain reasonably good constraints on the spectral index for four epochs between $2 - 15$~days post-FXT, ranging between $-1.9$ and $-0.4$. At 4 days post-FXT, we have a data point from \textit{Swift}/XRT in addition to those at optical frequencies, and so we construct a spectral energy distribution (see the bottom panel of Figure~\ref{fig:powerlaw_fits}). The purple shaded region in the bottom panel of Figure~\ref{fig:powerlaw_fits} denotes the photon index uncertainty on this \textit{Swift} data point. We fit a single power law to the optical and X-ray data points, and measure a spectral index of $-0.76 \pm 0.05$. When compared to the theoretical spectral index for optically thin synchrotron emission ($\frac{1 - p}{2}$), where $p$ typically lies in the range, $2 \lesssim p \lesssim 3$, we find that the two are consistent (we measure $p = 2.5 \pm 0.1$).

Given the power law-like light curves, the non-thermal spectrum of \thisAT, and its high optical luminosity ($M_r \sim -21$), it seems plausible that the optical emission at phases A and B originates from a GRB afterglow. To further explore this possibility, we compare the power law exponents from the optical light curves to those from analytical models in the literature \citep[\eg,][]{2002ApJ...568..820G, 2013NewAR..57..141G}.

First, we examine the decay rate measured in the first two days. We find that the initial decay rate is consistent across both bands ($\alpha_1 = -0.31 \pm 0.02$ in $r$-band, \mbox{$\alpha_1 = -0.33 \pm 0.05$} in $i$-band). We find that this observed decay rate is similar to a forward shock in a fast-cooling, optically thin regime \mbox{($\nu_{sa} < \nu_{c} < \nu_{obs} < \nu_{m}$):  $t^{-0.25}$} and is independent of the circum-burst density profile. Here, $\nu_{sa}$, $\nu_{c}$, $\nu_{obs}$ and $\nu_{m}$ are the self-absorption, cooling, observing and characteristic electron break frequencies, respectively. 

After the peak, we find that the decay rates change significantly depending on whether or not we include the late-time ($> 10$~days post-FXT) data. In the scenario where we include all the data, we find the second decay rate to be steeper; \mbox{$\alpha_3 = -1.4 \pm 0.4$, $-1.04 \pm 0.07$ and $-0.93 \pm 0.09$} at $r$-, $i$- and $z$-band, respectively. When combined with the measured spectral index ($p = 2.5 \pm 0.1$), we find that the optical emission after the light curve peak is most consistent with a forward shock adiabatically expanding into a homogeneous environment \citep[$\nu_{sa} < \nu_{m} < \nu_{obs} < \nu_{c}$;][]{2002ApJ...568..820G}. 

If we do not include the late time data points, we find that the post-peak decay rates are even steeper: \mbox{$\alpha_3 \approx -2.6 \pm 0.4$, $-2.2 \pm 0.4$ and $-2.5 \pm 0.1$} at $r$-, $i$- and $z$-band, respectively. These decay rates are consistent with $\alpha_3 = -p$, where $p$ is the same as that derived from the spectral index. In the GRB scenario this is interpreted as an optically thin, post-jet break decay rate \citep[$\nu_{sa} < \nu_{m} < \nu_{obs} < \nu_{c}$;][]{1999ApJ...519L..17S}. A jet break occurs when the jet has decelerated sufficiently such that the beaming cone encompasses the whole shock front, so the observer sees the intrinsic decay rate of the afterglow. We note that the steeper decay rates are consistent with those found by \citet{BrightEP240414a} with ATCA on timescales of tens of days. 

From the two decay rates ($\alpha_1$ and $\alpha_3$), we can infer that the forward shock transitions from a fast to slow cooling regime (\ie, radiative to adiabatic cooling) between 2 and 3 rest-frame days post-FXT. During the same time period, we observe a rapid rise in the optical light curve. In simple, canonical forward shock models, the steepest possible light curve rise is $\propto t^{7/4}$ and occurs in the regime $\nu_{m} < \nu_{obs} < \nu_{sa} < \nu_{c}$, which does not agree with our findings pre- and post-peak. 

Despite the lack of consistency with canonical GRB afterglow models, the observed variability is not uncommon in optical afterglows of GRBs. \cite{liang_comprehensive_2013} found a number of GRB afterglows that exhibit optical re-brightening on timescales of $\lesssim 1$\,day, and a previous Einstein Probe source (\EPxx{240315a}) displayed a significant optical and X-ray re-brightening $\approx 3$~days post-burst \citep{LiuEP240315a}. The most likely explanation is that the optical re-brightening is a consequence of refreshed shocks \citep{rees_refreshed_1998,sari_impulsive_2000} which result from slower shells ejected from the source after the main burst catching up with and re-energising the main afterglow shock at late times, as has been suggested for \GRBxx{030329} \citep{granot_nature_2003,moss_signature_2023}, \GRBxx{160821B} \citep{2019ApJ...883...48L}, and \GRBxx{170817A} \citep{lamb_grb_2020}. The dynamical time scale in the ejecta is comparable to the time of observation within a decelerating blastwave picture \citep{moss_signature_2023}. We note that the rebrightening episode during phase B starts at $t \sim 1.5$ rest-frame days with a rise time of $\Delta t \sim 1.5$ rest-frame days, thus satisfying $\Delta t / t \sim 1$. We find that the timescale post-FXT and duration of the flare from \thisAT\ is very similar to that observed by \cite{moss_signature_2023}. Other plausible mechanisms include external density fluctuations \citep[\eg,][]{dai_hydro_2002}, although this is difficult to reconcile with such a sharp flux increase \citep{nakar_piran_2003}, or possibly magnetar central engine activity \citep{gao_late_2015}. Unfortunately, we do not have comprehensive spectral and temporal coverage to rule out these other scenarios with confidence. 

\cite{BrightEP240414a} propose that the radio emission detected from \thisEP\ is most likely explained by a moderately relativistic outflow that is frequently detected in long GRBs. \cite{BrightEP240414a} further note that the only limit on gamma-ray emission at the time of \thisEP\ comes from Konus-Wind, and implies $L_{\rm iso} \lesssim 10^{51}$\,\ergs\ at our measured redshift \citep{Tsvetkova2017}. Hence, a low-luminosity GRB may have accompanied \thisEP, but remained undetected. 

\subsection{\thisEP\ in context with other FXTs}

\begin{figure*}
    \centering
    \includegraphics[width=\linewidth]{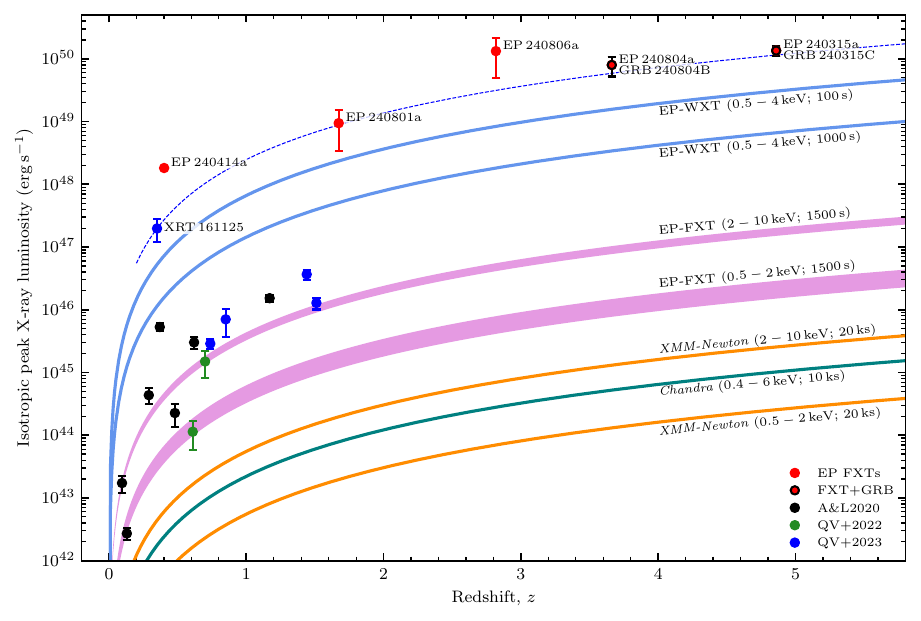}
    \caption{
        Isotropic peak X-ray luminosities of FXTs as a function of redshift. We compare the sample of EP FXTs with reported redshifts to the A\&L2020, QV+2022 and QV+2023 samples (see the main text for details). We highlight the isotropic peak luminosity that XRT\,161125 would have, assuming it has a redshift, $z \geq 0.2$. The simulated sensitivity curves are plotted for EP-WXT and EP-FXT, for the nominal exposure times listed \citep[taken from][]{Yuan2022_EinsteinProbe, Zhang2022_FXT_sensitivity}. We also plot sensitivity curves for the characteristic detection limits of \XMMNewton\ and \Chandra, taken from \cite{Watson2001_XMM-Newton_sensitivity} and the \Chandra\ Observatory guide (\url{https://cxc.harvard.edu/proposer/POG/html/chap6.html}).
    }
    \label{fig:FXT peak luminosity vs redshift}
\end{figure*}

EP has released GCNs for 27 FXT events, with eight optical counterparts discovered, and redshifts measured for just five of these.\footnote{As of 14 August 2024.} Generally, they appear to be significantly more distant than the redshifts inferred for the \XMMNewton\ \citep{alp_larssson_20} and \Chandra\ \citep{Quirola2022, Quirola2023} FXTs.
Here we qualitatively compare the properties of the various samples of FXT objects, to explore their similarities and differences.
Despite the properties and discovery methods of the \Chandra\ and \XMMNewton\ FXT samples being distinct from those uncovered by EP, a comparison of their properties can still provide useful insights, as presented below.

Including \thisEP, there are five FXT events from EP that have known redshifts; these include 
\EPxx{240315a} \citepalias[\mbox{$z = 4.859$};][]{Gillanders24, LevanEP240314a, LiuEP240315a},
\thisEP\ (\mbox{$z = 0.4018$}; see Section~\ref{SEC: Redshift estimation}),
\EPxx{240801a} \citep[\mbox{$z = 1.673$};][]{QuirolaVasquezGCN37013},
\EPxx{240804a} \citep[\mbox{$z = 3.662$};][]{2024GCN.37039....1B} and
\EPxx{240806a} \citep[\mbox{$z = 2.818$};][]{QuirolaVasquezGCN37087}.

To this set, we add XT\,030206, XT\,070618, XT\,100424, XT\,110621, XT\,151128, XT\,151219 and XT\,161028 extracted from the \XMMNewton\ catalogue presented by \cite{alp_larssson_20} (hereafter the A\&L2020 sample). From archival \Chandra\ observations spanning \mbox{$2000 - 2014$}, we include
XRT\,041230 and XRT\,080819 \citep[][hereafter the QV+2022 sample]{Quirola2022},
and from \Chandra\ observations spanning $2014 - 2022$, we include
XRT\,150322 \citep{Xue2019_XRT150322, Quirola2023},
XRT\,161125 \citep[][]{Quirola2023},
XRT\,170901 \citep{Lin2019ATel13171_XRT170901, Quirola2023},
XRT\,191223 \citep[][]{Quirola2023} and
XRT\,210423 \citep{Lin2021ATel14599_XRT210423, Quirola2023}, hereafter the QV+2023 sample.

In Figure~\ref{fig:FXT peak luminosity vs redshift}, we compare the EP FXTs to the A\&L2020, QV+2022 and QV+2023 samples \citep{alp_larssson_20, Quirola2022, Quirola2023}. For ease of comparison, we convert all recorded peak flux values to isotropic peak luminosity.

From Figure~\ref{fig:FXT peak luminosity vs redshift} we see that the historical samples of FXTs (\ie, the samples of A\&L2020, QV+2022, and QV+2023) all tend to have typical (isotropic-equivalent) luminosities of $\sim {\rm a \ few} \times 10^{42} - 10^{46}$\,\ergs, and redshifts, $z < 1.6$.\footnote{The exception to this is XRT\,161125, which we discuss later.} In comparison, the sample of (known-redshift) EP FXTs are significantly more luminous; they all have $L_{\rm iso} > 10^{48}$\,\ergs, and redshifts ranging from $z \approx 0.4 - 5$. Given that two of these, \EPxx{240315a}/\GRBxx{240315C} and \EPxx{240804a}/\GRBxx{240804B} (and seven of the full sample of 27), are likely the prompt X-ray emission from long GRBs \citep{Fredericks2024_GCN37071_EP240804a_GRB, LevanEP240314a, LiuEP240315a}, this luminosity separation is not unexpected given that we simply compute isotropic luminosity.\footnote{Assuming the low-redshift FXT samples are not beamed events.} \thisEP\ has lower luminosity than the high-redshift EP FXTs that are confirmed GRBs, and is at a comparable redshift to the A\&L2020, QV+2022 and QV+2023 samples. However, it seems unlikely to be related to the \Chandra\ and \XMMNewton\ sources as it is more than two orders of magnitude more luminous.\footnote{Aside from XRT\,161125.}

Given the EP-WXT observing strategy of \mbox{$t_{\rm exp} = 900 - 1200$\,s} \citep{Yuan2022_EinsteinProbe}, a source with $L_{\rm iso} = 10^{45}$\,erg\,s$^{-1}$ would only be detectable at a redshift, $z \lesssim 0.1$. Hence, the X-ray luminosity parameter space probed by EP is distinct from that which most previously known FXTs occupy. EP is designed such that it has a higher luminosity threshold than \XMMNewton\ and \Chandra, making it insensitive to the parameter space occupied by the historical FXT samples (see Figure~\ref{fig:FXT peak luminosity vs redshift} for a visualisation of the sensitivity limits of EP, \XMMNewton\ and \Chandra). Conversely, \XMMNewton\ and \Chandra\ have not detected FXTs that match those uncovered by EP; this is indicative of EP-detected events being intrinsically (volumetrically) rare events that have been routinely missed by the narrow fields of view of these observatories. However, the large field of view of EP has ensured their detection.
Thus, it appears that EP is primarily sensitive to the rare, luminous events that are broadly consistent with cosmological GRB events (out to high-$z$), whereas the deep, narrower field-of-view observatories are more suited to uncovering intrinsically fainter (but volumetrically more common) FXTs.

The one exception to this broad picture is XRT\,161125, which has $L_{\rm iso} = 2 \times 10^{47}$\,erg\,s$^{-1}$, and a redshift, $z = 0.35^{+0.05}_{-0.15}$ \citep{Quirola2023}. This redshift is based on an unconfirmed association with the marginally resolved $r = 24.2$ mag host galaxy in the DECam Legacy Surveys \citep{LegacySurveys}. Thus, the association is far from secure and it is plausible that this is merely a chance coincidence (\citealt{Quirola2023} estimate a chance association probability of $< 0.095$), with XRT\,161125 being at a much larger distance, making it similar in physical nature to the high-redshift EP transient sample. In Figure~\ref{fig:FXT peak luminosity vs redshift}, we illustrate where in the luminosity/redshift parameter space XRT\,161125 would lie for a range of redshifts \mbox{($z \geq 0.2$)}. We find that the isotropic peak luminosity inferred for XRT\,161125 closely resembles that of the four EP FXTs at $z \gtrsim 1.5$ (\EPxx{240315a}, \EPxx{240801a}, \EPxx{240804a} and \EPxx{240806a}).

Alternatively, if the host galaxy association for XRT\,161125 is correct, then we note that XRT\,161125 possesses a (projected) offset of $\approx 13.2$\,kpc \citep{Quirola2023}. This value is approximately half the separation observed here for \thisEP/\thisAT\ (27\,kpc; see Section~\ref{SEC: Redshift estimation}).

While \thisEP/\thisAT\ has a redshift in agreement with the historical sample of FXT events, its relatively high luminosity paired with our light curve analysis (see Section~\ref{sec:analysis}) indicates that this event is consistent with a GRB. In Figure~\ref{fig:FXT peak luminosity vs redshift}, we see that \thisEP\ is somewhat fainter than (but close to) the luminosity of the rest of the sample of known-redshift, EP-detected FXT events. We conclude that \thisEP\ likely belongs to the same population as these other EP-discovered FXTs, thus making it the most nearby (and intrinsically faintest) FXT event detected by EP linked to a GRB.

\section{Summary and conclusions} \label{sec:conclusions}

This Letter presents the discovery of \thisAT, the optical counterpart to the FXT \thisEP. The most likely host galaxy is identified as SDSS~J124601.99-094309.3 which we show has a redshift, \hostz, with \thisEP\ having a large (projected) radial offset of \hostPhysicalSep. We compute the chance coincidence of \thisEP\ with \hostGalName\ to be $\lesssim 0.01$.

Optical photometric follow-up of \thisAT\ reveals a luminous, multi-peaked light curve. Three distinct phases are identified: $\lesssim 2$, $2 - 10$ and $\gtrsim 10$~rest-frame days post-FXT (phases A, B and C, respectively).

Phase A rapidly fades from an initial discovery AB magnitude of $M_i \approx -20.4$ to $M_i \approx -19.5$ within 1.5 (rest-frame) days, before re-brightening on a $\sim 0.5$\,d timescale to $M_i \sim -21$. This second phase (B) then declines over a $\sim 10$~day period, before transitioning into the slowly evolving phase C. We find that phases A and B can be readily explained with a combination of simple power law fits, indicating the plausibility of the early light curve being powered by GRB afterglow radiation (with some re-brightening effects).

Although the rapid timescale and high luminosity ($M_r \sim -21$) of phase B is reminiscent of LFBOTs, we do not favour this scenario of thermal emission based on detailed analysis with \mosfit\ that yields inferred ejecta velocities of $0.8 \lesssim v_{\rm ej} / {\rm c} \lesssim 0.9$. The high X-ray luminosity, together with the red optical colours and achromatic evolution, support a non-thermal powering mechanism. An SED fit to the composite optical and \textit{Swift}/XRT data during phase B yields a spectral index of $0.76 \pm 0.05$, consistent with optically thin synchrotron emission. 

The third phase (C), starting $\sim 10$ rest-frame days post-FXT, shows evidence of an emerging SN component, the timescale and luminosity of which is consistent with broad-line SNe Ic, lending further support to a long GRB interpretation.

By comparing the full sample of known-redshift EP-discovered FXTs with the historical \XMMNewton\ and \Chandra\ samples of FXTs, we find that EP FXT events are consistently and systematically more luminous than all previously discovered FXTs. \thisEP\ is the most nearby of the EP-discovered FXT sample to date by a considerable margin and although its redshift closely resembles the \XMMNewton\ and \Chandra\ samples, it is significantly more luminous. EP is optimised for uncovering volumetrically rare, high-redshift, X-ray-luminous events, many of which appear to be linked to gamma-ray bursts.

\thisEP\ is likely X-ray emission from a low-luminosity GRB, with the early optical observations dominated by afterglow, followed by thermal emission from a GRB-SN. \thisEP\ is thus the lowest luminosity, and most nearby, EP-discovered FXT+GRB event uncovered to date.

\section*{Acknowledgments}

The authors thank the anonymous reviewer for detailed comments that improved the quality of the manuscript.

SJS, SS, KWS and DRY acknowledge funding from STFC Grants ST/Y001605/1, ST/X001253/1, ST/X006506/1 and ST/T000198/1. 
SJS acknowledges a Royal Society Research Professorship and the Hintze charitable foundation. LR acknowledges support from the Canada Excellence Research Chair in Transient Astrophysics (CERC-2022-00009). HFS is supported by Schmidt Sciences, LLC. MN is supported by the European Research Council (ERC) under the European Union's Horizon 2020 research and innovation programme (grant agreement No.~948381) and by UK Space Agency Grant No.~ST/Y000692/1. 
FC acknowledges support from the Royal Society through the Newton International Fellowship programme (NIF/R1/211296).
AJC acknowledges support from the Hintze Family Charitable Foundation. TWC \& AA acknowledge the Yushan Fellow Program by the Ministry of Education, Taiwan for the financial support (MOE-111-YSFMS-0008-001-P1). SY acknowledges the funding from the National Natural Science Foundation of China under Grant No. 12303046 and the Henan Province High-Level Talent International Training Program.
A.B is grateful to the Royal Society, United Kingdom and also acknowledges SERB (SB/SRS/2022-23/124/PS) for financial support.

Pan-STARRS is primarily funded to search for near-earth asteroids through NASA grants NNX08AR22G and NNX14AM74G. The Pan-STARRS science products for transient follow-up are made possible through the contributions of the University of Hawaii Institute for Astronomy and Queen's University Belfast.

This publication has made use of data collected at Lulin Observatory, partly supported by MoST grant 108-2112-M-008-001. We thank Lulin staff and NCU-GREAT team members for observations.

The Liverpool Telescope is operated on the island of La Palma by Liverpool John Moores University in the Spanish Observatorio del Roque de los Muchachos of the Instituto de Astrofisica de Canarias with financial support from the UK Science and Technology Facilities Council.

Based on observations obtained at the international Gemini Observatory (under program IDs GN-2024A-Q-128 and GS-2024A-DD-201), a program of NSF NOIRLab, which is managed by the Association of Universities for Research in Astronomy (AURA) under a cooperative agreement with the U.S. National Science Foundation on behalf of the Gemini Observatory partnership: the U.S. National Science Foundation (United States), National Research Council (Canada), Agencia Nacional de Investigaci\'{o}n y Desarrollo (Chile), Ministerio de Ciencia, Tecnolog\'{i}a e Innovaci\'{o}n (Argentina), Minist\'{e}rio da Ci\^{e}ncia, Tecnologia, Inova\c{c}\~{o}es e Comunica\c{c}\~{o}es (Brazil), and Korea Astronomy and Space Science Institute (Republic of Korea). This work was enabled by observations made from the Gemini North telescope, located within the Maunakea Science Reserve and adjacent to the summit of Maunakea. We are grateful for the privilege of observing the Universe from a place that is unique in both its astronomical quality and its cultural significance.

\facilities{
    \begin{itemize}
        \item LOT
        \item Pan-STARRS
        \item Gemini
    \end{itemize}
}

\software{
    \begin{itemize}
        \item \texttt{Astropy} \citep{2013A&A...558A..33A,2018AJ....156..123A,2022ApJ...935..167A}
        \item \texttt{Matplotlib} \citep{Hunter2007_matplotlib}
        \item \texttt{NumPy} \citep{Harris2020}
        \item \texttt{pandas} \citep{mckinney-proc-scipy-2010,reback2020pandas}
        \item \texttt{DRAGONS} \citep{Labrie2023_DRAGONS, DRAGONS_zenodo}
    \end{itemize}
}

\appendix \twocolumngrid

\section{Optical counterpart follow-up} \label{SEC: Appendix - Optical counterpart follow-up}

\subsection{Historical non-detections of an optical excess in ATLAS and Pan-STARRS} \label{SEC: Appendix - Optical counterpart follow-up - ATLAS and Pan-STARRS non-detections}

We checked the survey history of the Asteroid Terrestrial-impact Last Alert System \citep[ATLAS;][]{Tonry2018_ATLAS} at this sky position for any excess flux before the EP X-ray discovery, or for outbursts from the source that may suggest a Galactic flaring star in chance coincidence with the putative host galaxy. The ATLAS system is comprised of four 0.5-m telescopes that provide continuous sky coverage. ATLAS typically images the visible sky four times in 24\,hr when all four units are operating, and all data are processed rapidly to search for extragalactic transients \citep{Smith2020_ATLAS}.  The last observations of this position with ATLAS were $5 \times 30$\,s exposures obtained between MJDs 60402 and 60404 with the Sutherland unit in South Africa, with a stacked $3 \sigma$ forced upper limit of $o > 21.6$ measured from the difference images. There is no indication of any activity in the survey history of ATLAS. There is also no indication of a flux excess in forced photometry at this position in the history of Pan-STARRS survey data in 194 separate images between 17 March 2017 to 19 March 2024, to typical AB magnitudes of $w_{\rm PS} \gtrsim 22$.

\subsection{Optical photometric follow-up} \label{SEC: Appendix - Optical counterpart follow-up - Optical photometric follow-up}

After the initial discovery, we continued to obtain follow-up photometry of \thisAT\ with the LOT in $gri$-bands. We employed the Kinder pipeline \citep{kinderpip} to conduct PSF photometry for \thisAT\ with template subtraction. The derived magnitudes and upper limits were determined by calibrating against SDSS field stars in the AB system.  
 
Additionally, follow-up photometric observations with Pan-STARRS commenced on MJD 60416.41, 2.0 days after the discovery of \thisEP. Pan-STARRS (PS) comprises a twin 1.8-m telescope system (Pan-STARRS1 and Pan-STARRS2), both situated atop Haleakala mountain on the Hawaiian island of Maui \citep{Chambers2016arXiv_PanSTARRS1}. Images of \thisAT\ were obtained with Pan-STARRS1 (PS1), equipped with a 1.4~gigapixel camera and a 0.26~arcsec pixel scale, providing a $\sim 7$\,deg$^{2}$ field of view. The images were obtained in the $grizy$ filters \citep{Tonry2012}. Images were processed with the Image Processing Pipeline \citep{magnier2020a, waters2020}. The images were astrometrically and photometrically calibrated \citep{magnier2020c} and individual frames were co-added with median clipping to produce stacks, on which PSF photometry was performed \citep{magnier2020b}. 

Finally, we obtained $riz$-band images with the Gemini-North/GMOS-N instrument commencing on MJD 60489.28 ($\sim 75$\,d after the FXT trigger), under the program ID GN-2024A-Q-128 (PI:~M.~Huber). We also obtained images of the field in $i$-band with Gemini-South/GMOS-S starting MJD 60507.98 ($\sim 94$\,d post-FXT) under a Director's Discretionary Time proposal (PI: S. J. Smartt). These images were reduced using the \texttt{DRAGONS} pipeline \citep{Labrie2023_DRAGONS, DRAGONS_zenodo} and following standard recipes. The GMOS images are considerably deeper than reference images of the field from PS1, and image subtraction was therefore not possible. Although \thisAT\ is detected in the first set of GMOS images on MJD 60489 (see Figure~\ref{fig:AT204gsa detection images}, lower left panel), the transient is faint and host contamination is quite significant at the location. We used the \texttt{PSF} code \citep[][]{2023ApJ...954L..28N} to perform  photometry on the GMOS images without subtraction, and the estimated magnitudes are likely overestimated due to host galaxy contamination, particularly in the $z$-band.

The optical light curve is shown in Figure~\ref{fig:AT2024gsa_LC} and the summary of photometric observations is provided in Table~\ref{tab:Photometry}.

\subsection{Swift X-ray analysis} \label{SEC: Appendix - Optical counterpart follow-up - Swift X-ray analysis}


The EP-WXT detection was reported by \cite{EP240414aDiscovery} as a trigger starting \EPdiscoveryUTC\ (MJD~\EPdiscoveryMJD), with a peak flux of $\sim 3 \times 10^{-9}$\,\ergscm\ in the $0.5 - 4$\,keV band. This was followed up with the EP-FXT observation starting on 14 April 2024 at 11:50:01 UTC, approximately 2~hours after the EP-WXT detection \citep{EP240414aFXT}. The derived unabsorbed flux was reported by \cite{EP240414aFXT} to be $3.5 \pm 0.8 \times 10^{-13}$\,\ergscm\ in the $0.5 - 10$\,keV band; the source was seen to have rapidly faded.\footnote{\cite{EP240414aFXT} obtained this value by fitting the X-ray EP-FXT spectrum with an absorbed power law model, invoking \mbox{$N_{\rm H} = 3.35 \times 10^{20}$\,cm$^{-2}$} and a photon index of $1.7 \pm 0.3$.}

After the initial detection of \thisEP, three \textit{Swift}/XRT observations were performed in photon counting mode. The first observation made on 18 April 2024 detected a low number of photons and, therefore, did not allow us to invoke $\chi^2$ statistics. We grouped the spectra using the ftools task \texttt{grppha} (\textit{HEASoft} v6.33.2) to have at least one count per bin. Spectra were fitted using \textsc{xspec 12.14.0h} \citep{Arnaud1996}. Using W-statistics and background-subtracted Cash statistics \citep{Wachter1979}, we fit the spectra using an absorbed power law model. Owing to limited statistics, it was difficult to constrain $N_{\rm H}$ and the photon index. Therefore, we fixed these values to that reported by \cite{EP240414aFXT}. The obtained X-ray flux (\mbox{$5 \pm 2 \times 10^{-13}$\,\ergscm}; $0.3 - 10$\,keV) was found to be consistent with that reported by the EP team. Due to the rapid decay of the source in X-rays, it was not possible to fit the spectra of the other two observations made on  24 and 27 April 2024.~Therefore, we combined data of these two observations and computed a 95{\%} confidence level upper limit count rate using the prescription given by \citep{Gehrels86}; the corresponding flux value was found to be $2.25 \times 10^{-13}$\,\ergscm\ using the \textsc{webpimms heasarc} tool, assuming a photon index~of~2.

\section{Color evolution of \thisAT}\label{sec:color_evol}

Here we include a visual demonstration of the intrinsically red color of \thisAT, compared with \ATxx{2018cow} \citep{2018ApJ...865L...3P}, \GRBxx{071010A} \citep{2008MNRAS.388..347C} and \SNxx{1998bw} \citep{2011AJ....141..163C}. It is apparent that \thisAT\ is significantly redder than the LFBOT \ATxx{2018cow} and also \SNxx{1998bw} at later times. The early color evolution of \thisAT\ is somewhat similar to \SNxx{1998bw}, however the luminosity is much higher relative to \SNxx{1998bw} at a similar epoch (see Section~\ref{sec:analysis} for more details).

\begin{figure*}
    \centering
    \includegraphics[width=0.8\linewidth]{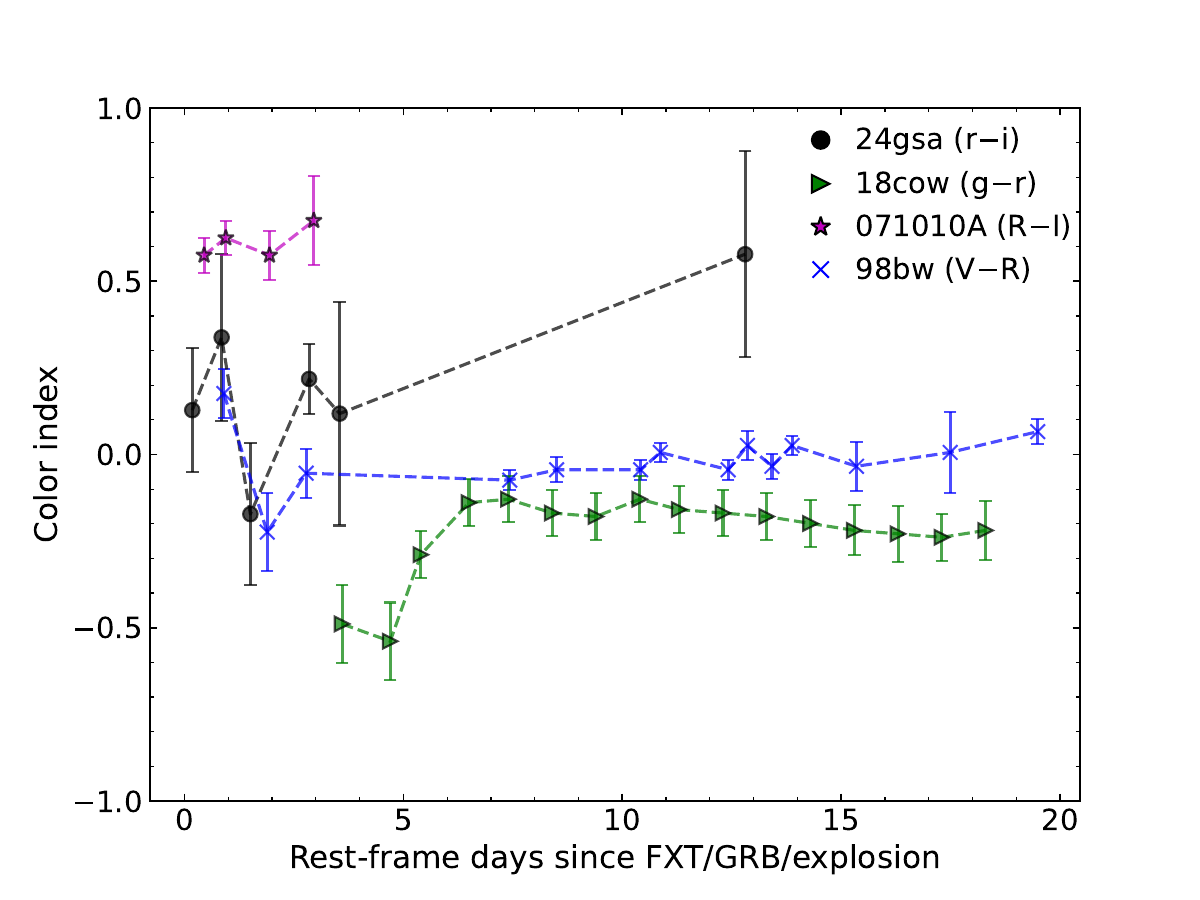}
    \caption{The ($r-i$) color evolution of \thisAT\ at early times (phases A and B), campared to the ($g-r$) color evolution of \ATxx{2018cow} \citep{2018ApJ...865L...3P}, ($R-I$) color evolution of \GRBxx{071010A} \citep{2008MNRAS.388..347C}, and ($V-R$) color evolution of \SNxx{1998bw} \citep{2011AJ....141..163C}.
    }
    \label{fig:color}
\end{figure*}


\begin{table*}
    \renewcommand*{\arraystretch}{1.1}
    \centering
    \caption{
        Optical photometry of \thisAT. Reported magnitudes are not corrected for the foreground Galactic extinction of \mbox{$E (B - V) = 0.033$} AB mag \citep{Schlafly2011}. The errors for the optical photometry are quoted to $1 \sigma$, and upper limits are $3 \sigma$. $T_0$, the time of trigger for \thisEP, is MJD~60414.40986. 
    }
    \begin{tabular}{ccccccc}
        \toprule
        
        MJD       &$T_{\rm mid} - T_0$      &Telescope    &Filter    &Total exposure    &Apparent magnitude    &Error       \\
                  &(observer-frame days)    &             &          &time (s)          &(AB mag)              &(AB mag)    \\

        \midrule

        60414.58    &0.17     &LOT             &$r$    &3300    &21.55       &0.08    \\
        60414.60    &0.19     &LOT             &$i$    &3600    &21.40       &0.16    \\
        60414.75    &0.34     &LOT             &$g$    &1800    &21.90       &0.12    \\
        60415.61    &1.20     &LOT             &$i$    &1500    &21.69       &0.23    \\
        60415.64    &1.23     &LOT             &$r$    &5700    &22.05       &0.07    \\
        60416.40    &1.99     &PS              &$r$    &900     &$>22.40$    &$-$     \\
        60416.41    &2.00     &PS              &$i$    &900     &22.37       &0.19    \\
        60416.42    &2.01     &PS              &$z$    &900     &21.95       &0.20    \\
        60416.56    &2.15     &LOT             &$r$    &6000    &22.05       &0.14    \\
        60416.64    &2.23     &LOT             &$i$    &5400    &22.20       &0.15    \\
        60417.38    &2.97     &PS              &$i$    &1200    &20.86       &0.05    \\
        60417.40    &2.99     &PS              &$z$    &1200    &20.56       &0.06    \\
        60417.51    &3.10     &LOT             &$r$    &3300    &21.06       &0.15    \\
        60418.39    &3.98     &PS              &$z$    &600     &20.30       &0.06    \\
        60418.39    &3.98     &PS              &$i$    &600     &20.51       &0.06    \\
        60418.40    &3.99     &PS              &$r$    &600     &20.75       &0.08    \\
        60418.41    &4.00     &PS              &$g$    &600     &21.08       &0.12    \\
        60419.36    &4.95     &PS              &$i$    &400     &20.97       &0.09    \\
        60419.37    &4.96     &PS              &$y$    &200     &20.55       &0.27    \\
        60419.37    &4.96     &PS              &$z$    &400     &20.77       &0.09    \\
        60419.38    &4.97     &PS              &$r$    &480     &21.11       &0.31    \\
        60419.39    &4.98     &PS              &$g$    &480     &$>20.57$    &$-$     \\
        60420.37    &5.96     &PS              &$i$    &900     &21.30       &0.27    \\
        60420.38    &5.97     &PS              &$z$    &900     &21.31       &0.24    \\
        60432.36    &17.95    &PS              &$r$    &900     &22.89       &0.21    \\
        60432.37    &17.96    &PS              &$i$    &900     &22.29       &0.21    \\
        60432.38    &17.97    &PS              &$z$    &900     &22.32       &0.26    \\
        60436.37    &21.96    &PS              &$r$    &800     &$>22.49$    &$-$     \\
        60436.38    &21.97    &PS              &$i$    &800     &22.16       &0.31    \\
        60436.39    &21.98    &PS              &$z$    &1200    &22.02       &0.31    \\
        60455.28    &40.87    &PS              &$i$    &1600    &23.16       &0.29    \\
        60455.30    &40.89    &PS              &$z$    &1600    &22.63       &0.22    \\
        60459.28    &44.87    &PS              &$i$    &1600    &23.35       &0.35    \\
        60459.30    &44.89    &PS              &$z$    &1600    &22.65       &0.21    \\
        60463.28    &48.87    &PS              &$i$    &1600    &23.39       &0.33    \\
        60463.30    &48.89    &PS              &$z$    &1600    &22.67       &0.25    \\
        60467.28    &52.87    &PS              &$i$    &1600    &23.36       &0.31    \\
        60467.30    &52.89    &PS              &$z$    &1600    &23.14       &0.32    \\
        60489.28    &74.87    &Gemini-North    &$r$    &1260    &24.28       &0.27    \\
        60489.30    &74.89    &Gemini-North    &$i$    &1260    &23.76       &0.28    \\
        60489.32    &74.91    &Gemini-North    &$z$    &1080    &23.05       &0.27    \\
        60507.98    &93.57    &Gemini-South    &$i$    &870     &$>24.58$    &$-$     \\
        
        \bottomrule

    \end{tabular}
    \label{tab:Photometry}
\end{table*}

\bibliography{References}
\bibliographystyle{aasjournal}

\end{document}